\documentstyle[12pt,epsfig]{article}
\begin{document}
\begin{titlepage}
\begin{center}
\vspace*{3cm}

{
\Large  \bf   Tilted plateau in nuclear collisions:\\ data, models and MC-s
}
\vspace{2cm}

\begin{author}
\Large
K. Fia{\l}kowski\footnote{e-mail address: uffialko@th.if.uj.edu.pl},
R. Wit\footnote{e-mail address: wit@th.if.uj.edu.pl}

\end{author}

\vspace{1cm}

{\sl M. Smoluchowski Institute of Physics\\ Jagellonian University \\

30-059 Krak{\'o}w, ul.Reymonta 4, Poland}

\vspace{3cm}

\begin{abstract}
The linear dependence  of the particle spectra on rapidity is seen in the central region for asymmetric
heavy ion collision in the data and in the Monte Carlo results, similarly as 
in the fragmentation region for hadronic and ion collisions. The origin of such a behaviour  
is discussed. It is shown that the color string models
produce naturally such a shape if string ends are randomly distributed in rapidity.

\end{abstract}

\end{center}
\vspace{1cm}

PACS:  \\


\vspace{1cm}

\noindent

 23 December, 2004 \\

\end{titlepage}

\section{Introduction}

\par
   Recently, there has been a renewed interest in the energy dependence of multiplicity of hadrons from
   the multiparticle production processes. In particular,  the possibility of understanding
   some regularities observed in data as the consequence of simple assumptions on the production mechanism
   was considered.
\par
   In this note we discuss the shape energy dependence of the distributions in CM rapidity

 $$y=\frac{1}{2}\ln \frac{E+p_L}{E-p_L}$$

\noindent
and CM  pseudorapidity
$$\eta=-\ln~ tg (\theta/2)$$

\noindent
for charged pions and other hadrons produced in the nuclear, $p$-$p$ and $e^+e^-$ collisions.
 We present distributions obtained from the
 Monte Carlo generators and compare them with some data
and simple rules expected from various assumptions. Some remarks on the possible generality and
universality of production mechanism are given.
\par
   First, let us remind  that the rapidity distributions have many advantages as the tools for the
   discussion of the energy dependence of multiparticle production providing more details than just
   the average multiplicity. The shape of rapidity distribution is invariant under the Lorentz
   transformation along the longitudinal axis and changes rather slowly with ener\-gy. This was
   the original motivation for the Feynman hypothesis of a "plateau" in rapidity with an energy
   independent height (based on an analogy with electromagnetic bremsstrahlung). Since the integral
   of a single-particle inclusive distribution is equal to the average multiplicity, the "plateau"
   hypothesis leads to the logarithmic increase of average multiplicity. Similarly, the generalization
   of Feynman hypothesis to the multiparticle distributions leads to the KNO scaling for the
   multiplicity distributions.
   \par
  Today we know that Feynman hypothesis was wrong. There is no "rapidity plateau" of a fixed height;
  the rapidity distributions resemble rather Gaussian curves with both height and width increasing
  with energy. The average multiplicity grows with energy faster than logarithmically, and KNO scaling
  is also not exact. Still, there seem to be other simple regularities in the rapidity distributions.
  \par
  The rapidity distributions for asymmetric nuclear collisions are particularly in\-teres\-ting. Naive expectations
  were those of "two plateaus" differing by height and smoothly connected for particles which are
  slow in the CM frame.
  In the data, however, we see instead a complicated structure resulting in the "tilted plateau" (i.e., linearly
  decreasing function of rapidity) for the ratio of $d$-$Au$ to the $p$-$p$ data.
  We discuss here in more detail this and some other regularities seen in the recent data on $d$-$Au$ collisions
  {\cite{WB}, \cite{PHOB}}, comment on their interpretation within the wounded nucleon model {\cite{BC}}
  and compare them with the predictions of the default version of the  FRITIOF Monte Carlo generator
  \cite{FRIT}, \cite{FRIT2}.
   \par
   Before starting this discussion we analyze another rapidity range in which a similar linear decrease of
   spectra in rapidity was observed: the fragmentation region in hadron-hadron and nuclear collisions.
   It is well known that in this case
   another Feynman hypothesis works well: scaling (i.e. the energy independence) of the distribution
   in "Feynman $x$ variable", $x=p_L /p_L^{max}$ , which is approximately valid for $x\neq0$ (more precisely,
   for $x$ significantly bigger than $m/\sqrt{s}$ ). It is easy to check that it means  energy
   independence of the distribution in $y^{max} - y$ for moderate values of this variable. This does not
   determine the shape, nor does it answer the simple question: is the range of scaling in $y^{max} - y$ 
   (or $y-y^{min}$)
   energy independent, or does it extend logarithmically with energy, as the full range in rapidity does?
   In this note we address these questions for the data \cite {UA5}, for a simple model \cite{BJ}
   and for the events modelled by PYTHIA  \cite{SJO} and FRITIOF \cite{FRIT}, \cite{FRIT2} generators.
   As mentioned above, in the fragmentation region of $p-p$ collisions a similar 
    linear dependence on rapidity is observed
   as in the central region of asymmetric nuclear collision. We comment on the origin of this
   similarity in the models underlying Monte Carlo generators.

   \section{Fragmentation in the $p$-$p$ and heavy ion collisions}
\par
\begin{figure}[h]
\centerline{\epsfig{figure=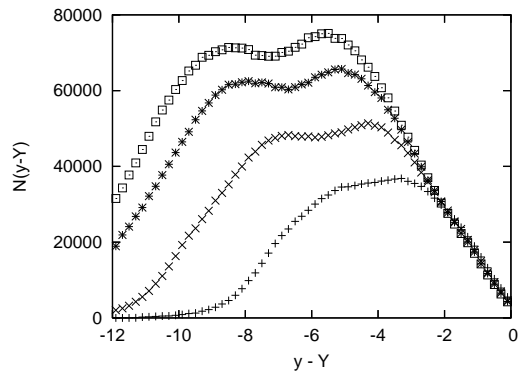, height =5cm}}
\caption{\label{fig1} {\small \sl PYTHIA results for the charged pion distributions in the
$p$-$p$ collisions at CM energies of $50, 200, 550$ and $900 ~ GeV$ (crosses, x-s, stars and squares, respectively).} }
\end{figure}
\begin{figure}[h]
\centerline{\epsfig{figure=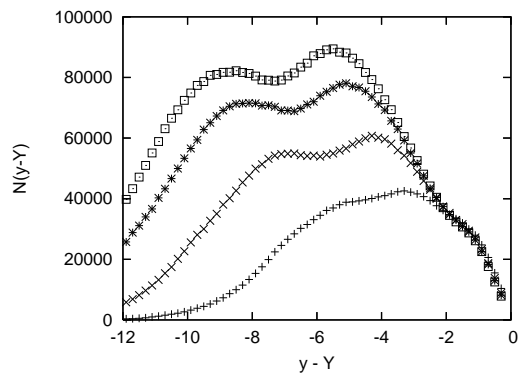, height =5cm}}
\caption{\label{fig2} {\small \sl PYTHIA results for the charged pions, kaons protons and antiprotons
 distributions in the $p$-$p$ collisions at CM energies of $50, 200, 550$ and $900 ~ GeV$
(crosses, x-s, stars and squares, respectively).} }
\end{figure}

   In Fig.\ref{fig1} and Fig.\ref{fig2} we show the PYTHIA results for the $p$-$p$ collisions at CM
   energies of $50, 200, 550$ and $900 ~ GeV$.

   The horizontal scale is $\eta'=\eta-y^{beam}$; thus all the consecutive plots are shifted so as to assure that
   the scaling corresponds to the energy independence of the right-hand end of the distribution. In Fig.\ref{fig1}
   only the charged pions are counted, whereas in Fig.\ref{fig2} charged kaons, protons and antiprotons are also
   included. We see
 that there is an approximate scaling, and the range of scaling in $\eta'$ increases with energy, but not as
   fast as the full pseudorapidity range; the range of the "plateau" increases as well.
   It is interesting to note that by counting only pions we get  almost exactly linear behaviour of spectra in
   the "scaling range", whereas for the "all charged" selection the points are less regularly distributed.
   \par
\begin{figure}[h]
\centerline{\epsfig{figure=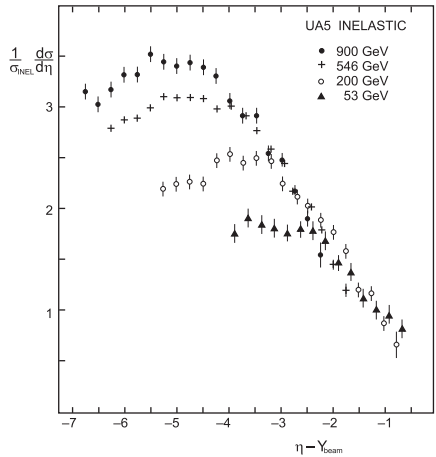, height =6cm}}
\caption{\label{fig3} {\small \sl ISR and UA5 results \cite{UA5} for the charged hadron
 distributions at CM energies of $50, 200, 550$ and $900 ~ GeV$
(triangles, circles, crosses and dots, respectively).} }
\end{figure}

   In Fig.\ref{fig3} the ISR and collider UA5 data \cite{UA5} at similar energies
   ($53, 200, 546$ and $900 ~GeV$) are shown in the
   same variable $\eta'$. We do not compare them directly with MC results, since the involved procedure of removing
   the UA5 detector effects is difficult to analyze. However, detailed inspection shows that the scaling
   seems to be broken more significantly than in Fig.1. The range covered by data depends on energy, but the
   slope with which the distribution falls at $\eta'\rightarrow 0$  increases  with increasing energy,
   which was not seen for MC results.

   \par
\begin{figure}[h]
\centerline{\epsfig{figure=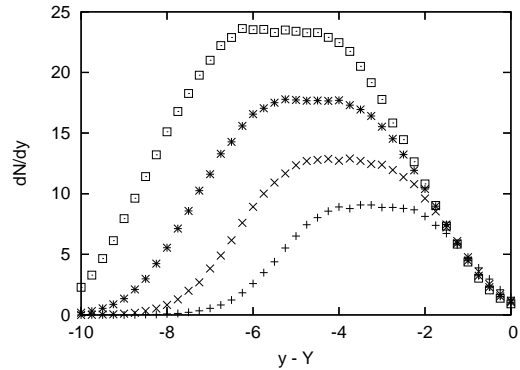, height =5cm}}
\caption{\label{fig4} {\small \sl FRITIOF results for the charged pions
 distributions in the $S$-$S$ collisions at $N$-$N$  CM energies of $20, 40, 80$ and $160 ~ GeV$
(crosses, x-s, stars and squares, respectively).} }
\end{figure}

   Similar approximate scaling is observed in the rapidity spectra from heavy ion collisions. As an example of the results
   from the MC calculations for such collisions we show in Fig.\ref{fig4} the distributions obtained from the FRITIOF
   generator for the central $S$-$S$ collisions. The sample was defined by requiring more than
   20 participants (out of 32
   nucleons) for the forward going ion. This condition is satisfied by 11-13\% of events; the percentage is slightly
   increasing with
   energy. We see that there is an approximate scaling for the range increasing with energy, and the shape of the
   distribution in this range is again linear.

\par
\begin{figure}[h]
\centerline{\epsfig{figure=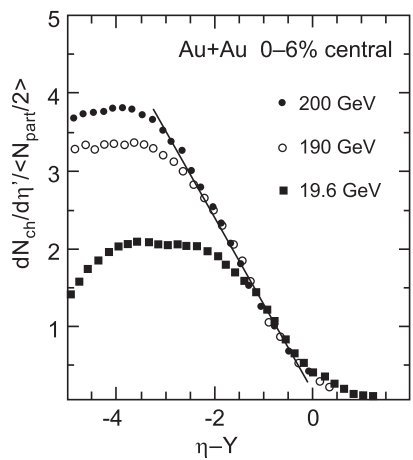, height =6cm}}
\caption{\label{fig5} {\small \sl PHOBOS results for the charged hadron
 distributions in the $Au$-$Au$ collisions at $N$-$N$ CM energies of $19.6, 130$ and $200 ~ GeV$
(squares, circles and dots, respectively).} }
\end{figure}

The recent data concerning rapidity distributions from the heavy ion collisions were presented by PHOBOS collaboration
\cite{PHOB}. The data collected at three energies for the most central $Pb$-$Pb$ collisions (6\% of the full sample
of events) are shown in Fig.\ref{fig5}. Their behavior is qualitatively identical to that of MC results from Fig.\ref{fig4}.

\par
It is interesting to note that an approximate scaling, a linear dependence of spectra on rapidity in the fragmentation
region and the increase of the range of such behavior with energy are not seen in the simplest production process: $e^+e^-$
annihilation into hadrons. Obviously, there are many problems when comparing this process with hadron- or heavy ion
collisions. To define rapidity, one needs to know the collision axis. It may be defined as the sphericity or thrust axis,
but the estimate of the direction of this axis  for some events  may be poorly determined.
The alternative is to define a sort of "energetical rapidity" independent on the direction of particle momentum. In both cases
we tried unsuccessfully to establish scaling in the MC results for the fragmentation region, even for the restricted range
of energies. One may attribute it to the thresholds for heavy quark production and/or energy dependence of the multijet
fraction of events. This should be kept in mind when discussing the possible interpretation of the effect.
\par
   The approximate scaling with the linear dependence on rapidity for the hadron-hadron and heavy ion collisions
   was recently explained by a simple model motivated by the nonabelian bremsstrahlung effect \cite {BJ}.
   \par
   In this model the first stage of the hadroproduction process consists of consecutive color exchanges
   between the pairs of partons from two colliding hadrons. The created color charges emit the final state hadrons
   by the bremsstrahlung process (uniformly in rapidity). If the initial partons are also uniformly
   distributed in rapidity (which corresponds to the
   $1/x$ spectrum in Feynman $x$ variable), the resulting distributions in the fragmentation region fall linearly with
   the rapidity variable $y -Y_{max}$, as seen in the data.
   \par
   The limit  of the linearity regime is determined by the condition that interacting partons must live longer than for
   some fixed time $\tau_0$, necessary to complete the color exchange. Then the linear increase of the spectrum
   stops at some rapidity $y_0$, depending on $\tau_0$ and on the parton transverse mass. In the central region a plateau
   in rapidity appears. Obviously, this picture is
   frame dependent; assuming arbitrarily that the assumptions are fulfilled in the CM frame (which seems to agree
   with the data) one breaks explicitly the boost invariance.

   \par
   Therefore one may prefer another formulation of the model, in which the bremsstrahl\-ung process is replaced
   by the breaking of color strings spanned between a parton from the projectile and a parton from the target. In this
   case the plateau in the central region appears as a result of the assumption of minimal energy of a string,
   corresponding to the minimal difference in rapidities of the partons at the ends of the string. There are
   no very short strings (slow in the CM frame), which would result in the linear increase of spectra down to the
   CM rapidities close to zero.

   \par
   The linear decrease in the fragmentation region may be easily understood if we write down the spectrum as the
   result of an integral over the string end distribution. For small $y_{max} - y$ we get

   $$\rho (y) = \int \overline{n}_{s} \kappa \theta (z-y) \rho_{s} (z)dz $$

   \noindent
   where $\overline{n}_s$ is the average number of contributing strings, $\rho_s(z)$ the distribution of the
   string end, and $\kappa \theta(z-y)$ is the distribution of hadrons from a single string.
   If $\rho_s(z)=const$, one gets linear $\rho(y)$.
   \par
   It appears that the presented picture approximates well the models underlying the Monte Carlo generators
   used in the previous section. In the PYTHIA generator there are indeed many strings formed in each hadron collision,
   and the distribution of the ends of most of them obeys the $1/x$ distribution. Thus it is not surprising that the
   results from this generator (shown, eg., in Fig.\ref{fig1}) agree reasonably well with the data shown in Fig.\ref{fig3}.
   For the heavy ion  collisions at moderate  energies the FRITIOF generator uses just two strings for each nucleon
   collission. However, in this generator the distribution from a single string is non-flat for the major part of rapidity
   range, and a superposition of strings from different nucleons appears to produce approximate
   linearity of the distribution  in this case as well.

 \section{Asymmetric nuclear collisions}

 \par
\begin{figure}[h]
\centerline{\epsfig{figure=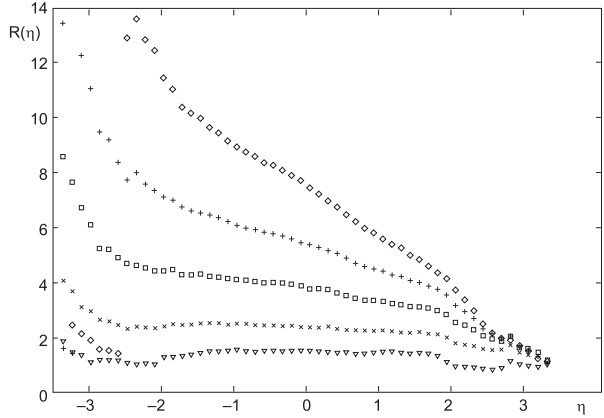, height =5.5cm}}
\caption{\label{fig6} {\small \sl PHOBOS results \cite{WB},\cite{PHOB} for the ratio $R(y)$ as a function of pseudorapidity
for 5 centrality bins (20\% each), ranging
from most peripheral (triangles) to the most central ones (diamonds).} }
\end{figure}

Recently, the PHOBOS collaboration presented new data on $d$-$Au$ collisions at RHIC energy \cite{WB}, \cite{PHOB}.
The data are divided into five equally populated centrality bins, ranging from the most central to the most
peripheral events. Considering the ratios of the rapidity spectra for these five samples to the $p$-$p$ data
$$R(y)=\frac{dN^{d-Au}/dy}{dN^{p-p}/dy}$$
 one observes the simple pattern, as shown in Fig.\ref{fig6}.

For a rather wide range of rapidities the ratio decreases linearly, and the slope is determined by the average number of
participants ("wounded nucleons") in the $Au$ nucleus for the given centrality range $w_{Au}^{(c)}$.
The deviations from the
linear dependence are seen for large negative rapidities (i.e., in the $Au$ nucleus fragmentation region,
where the nuclear cascade effects are needed to describe the data), and near the kinematical limits.
\begin{figure}[h]
\centerline{\epsfig{figure=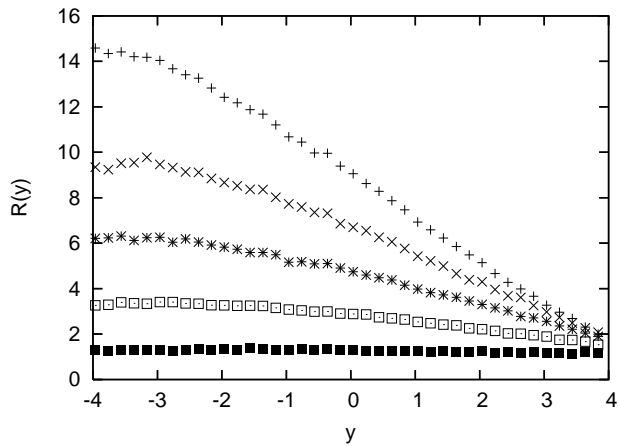, height =6cm}}
\caption{\label{fig7} {\small \sl FRITIOF MC results for the ratio $R(y)$ as a function of pseudorapidity
for 5 centrality bins (about 20\% each), ranging
from most peripheral (black squares) to the most central ones (crosses).}  }
\end{figure}
\par
We have checked that qualitatively similar results (Fig.\ref{fig7}) come out from the
FRITIOF generator, where
the color strings are assumed to form from each "wounded" nucleon. Precisely speaking, not all the strings in this model
are identical: the strings from
the "multiply wounded" nucleons are slightly longer and produce more particles, but this is a "second order correction".
The origin of the linear dependence on rapidity in this model is thus quite obvious: the "slow ends" of strings are
uniformly distributed in rapidity ($1/x$ distribution in Feynman $x$ variable), the distribution from a single string
is approximately flat and the number of strings in each CM hemisphere is just the number of wounded nucleons. In fact,
the linearity is much better for the "$d$-$Au$"/"$p$-$p$" ratio, than for the rapidity spectra, as the "non-flatness" effects
for single string in the $d$-$Au$ and $p$-$p$ collisions approximately cancel in the ratio.
\par
We did not compare directly the data  shown in Fig. \ref{fig6} and the MC results shown in Fig. \ref{fig7}. One
of the reasons is the absence of the nuclear cascade in the FRITIOF MC. Thus for $y < -2$ the data curve upwards,
and the MC downwards. Also the centrality bins were not exactly the same for both figures. However, we  see
clearly that the MC results reproduce well the "tilted plateau" with the slope proportional to the number of
wounded participants.
\par
The PHOBOS data were recently analyzed within the framework of the
wounded nucleon model \cite{BC}. In this model the symmetric and
antisymmetric components of the particle density
$$ G^{\pm}(\eta ) = \frac{dN(\eta)}{d\eta}\pm
\frac{dN(-\eta)}{d\eta}$$
\noindent
are proportional to the average symmetric and antisymmetric single nucleon contributions $<\Phi^\pm(\eta)>$,
which may be reconstructed from the data
$$<\Phi^\pm(\eta)> = \frac{\Sigma_c G^{(c)\pm}(\eta)}{\Sigma_c[w^{(c)}_{Au}\pm w^{(c)}_{d}]/2}$$.

\begin{figure}[h]
\centerline{\epsfig{figure=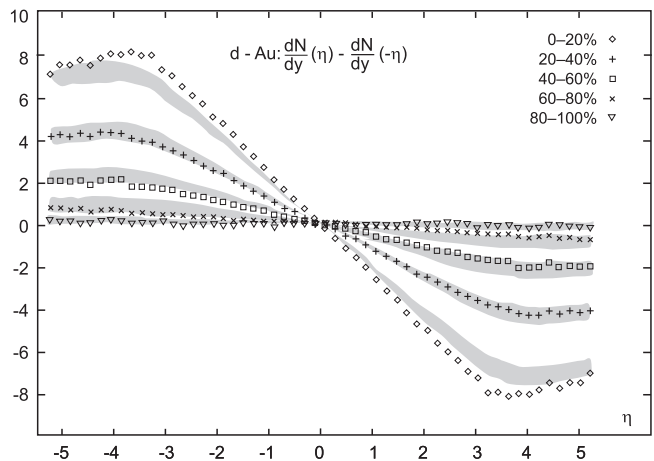, height =6cm}}
\caption{\label{fig8} {\small \sl The antisymmetric part of the $d$-$Au$ inclusive spectra  in pseudorapidity from PHOBOS
compared with the predictions of wounded nucleon model for five centrality bins \cite{BC}.} }
\end{figure}

For each centrality we get
$$G^{(c)\pm} (\eta) = \frac{w^{(c)}_{Au}\pm w^{(c)}_{d}}{2 }<\Phi^\pm(\eta)>.$$

\begin{figure}[h]
\centerline{\epsfig{figure=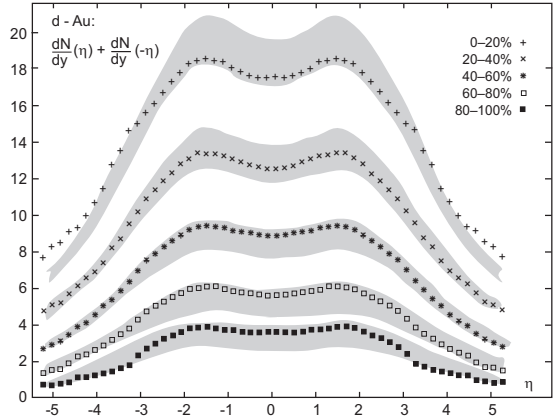, height =5.5cm}}
\caption{\label{fig9} {\small \sl The symmetric part of the $d$-$Au$ inclusive spectra  in pseudorapidity from PHOBOS
compared with the predictions of wounded nucleon model for five centrality bins \cite{BC}. } }
\end{figure}

In Fig.\ref{fig8} and Fig.\ref{fig9} \cite{BC} the model predictions are shown as grey bands and compared
with data for five centrality bins. For the antisymmetric components there is a qualitative
agreement, although the centrality dependence is underestimated: for peripheral events the data are closer
to the axis than the predictions, and for central events they are further from the axis. For the
symmetric component the agreement is better, although there appears a similar trend.

\par
\begin{figure}[h]
\centerline{\epsfig{figure=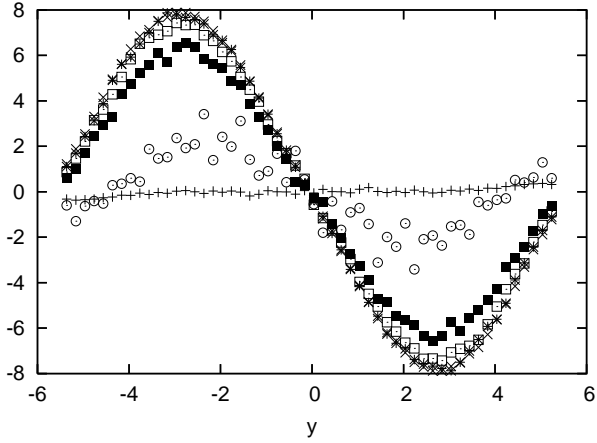, height =6cm}}
\caption{\label{fig10} {\small \sl The antisymmetric part of pseudorapidity spectra for $d$-$Au$ collisions from the
FRITIOF generator for five centrality bins, scaled by the difference $w_{Au} - w_d$.} }
\end{figure}
\begin{figure}[h]
\centerline{\epsfig{figure=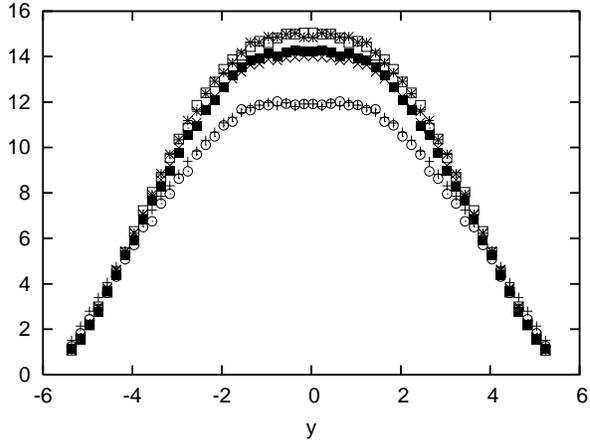, height =6cm}}
\caption{\label{fig11} {\small \sl The symmetric part of pseudorapidity spectra for $d$-$Au$ collisions from the
FRITIOF generator for five centrality bins, scaled by the sum $w_{Au} + w_d$.} }
\end{figure}

The average over centralities $<\Phi^\pm(\eta)> $ is meaningful only if separate contributions from
different centralities do not differ to much. We checked that for the FRITIOF MC the scaling of
$G^{(c)\pm}/(w^{(c)}_{Au}\pm w^{(c)}_{d})$ is approximately valid -- the deviations for most peripheral events are
due the improper estimate of $w_d$ in our sample. This is shown in Fig.\ref{fig10} and Fig.\ref{fig11}, where
the $p$-$p$ data are shown for comparison (crosses).

\par
We see that the FRITIOF generator (based on the string picture) gives qualitatively the same results
as the wounded nucleon model. One should stress that the contributions from a wounded nucleon as
extracted from the data \cite{BC} extend quite far in the "wrong" CM hemisphere. The same apparently happens
for single string contributions in FRITIOF. The flat distribution of the string ends extends far in both
hemispheres.
\par
The "{}tilted plateau"{} in the symmetric nuclear collisions is easy to understand in the string models.
Let us assume that there are $w_A$ strings from the wounded nucleons in nucleus $A$ and $w_B$ strings from
nucleus $B$, each string
producing flat rapidity spectrum and the "{}slow"{} string ends (in CM) are randomly distributed in rapidity
within the range from $-\Delta$ do $+\Delta$. We get (for $y$ in the same range)
$$
\rho (y) = \int \kappa[w_A \theta (z_1-y)  + w_b\theta (y-z_2)] \frac{dz_1dz_2}{4\Delta^2} =
\kappa[w_b+(w_A-w_b)\frac{\Delta -y}{2\Delta}]
  $$
  and
  $$ R(y) =[w_b+(w_A-w_b)\frac{\Delta -y}{2\Delta}]/2
  $$
Note that the two integrations are independent and one may use a single integration variable $z$ instead of two,
apparently assuming that the strings from both sides always merge.
This description is a fair approximation to the data (Fig.6) and to the results of FRITIOF (Fig.7),
when $\Delta \approx 3$ and the experimental values of $w_A$ and $w_B$ for each centrality are used.
As noted above, this suggests the strings related to each nucleus extend far into the "{}wrong"{} CM hemisphere.

\section{Conclusions and outlook}
\par
We have discussed rapidity spectra from the hadron-hadron and nuclear collisions.
The recent observation of linearly decreasing spectra in the fragmentation region and its interpretation
in a simple model based on the idea of nonabelian bremsstrahlung were recalled and the string version of
this model was shown to account for the successes of Monte Carlo generators in the  data description.
The main subject of our discussion were, however, the rapidity distributions in the central region from the
asymmetric heavy ion collisions. Reinterpreting the successes of wounded nucleon model in the language of
string model we have shown that the origin of "tilted plateau" in this case is the same as that of the linear decrease
in the fragmentation region: superposition of many "mini-plateaus" from single strings with a uniform distribution
of the string ends. Again, this accounts for the successes of Monte Carlo calculations using generators with the
appropriate description of string generation and decay.
\par
It is interesting to note that the scaling in the fragmentation region with the li\-near\-ly decreasing rapidity spectra
is not observed in the simplest hadroproduction process: $e^+e^-$ annihilation. At the first sight it seems to
contradict the idea of "non-abelian bremsstrahlung mechanism", which should be observed in the purest form for the
$q \overline q$ final state. However, in the string picture the linear decrease results from the uniform string
fragmentation and from  the uniform string end distribution. The first is true for the $q \overline q$ final state,
but the
second condition is not fulfilled. The detailed investigation of the shape and energy dependence of the rapidity
spectra from the $e^+e^-$ annihilation would be valuable, especially with the separation of the final state quark
flavors and two jet events.\\

{\bf Acknowledgements}
\par
We would like to thank  A. Bia{\l}as for suggesting this investigation and A. Kota{\'n}ski for reading the manuscript. 
One of us (KF) is grateful for
a partial financial support by the KBN grant \# 2 P03B 093 22, and  RW for
a partial financial support by the KBN grant \# 2 P03B 096 22.

\end{document}